\documentclass[useAMS,usenatbib]{mn2e}

\usepackage{amsmath}
\usepackage{graphics,epsfig,graphicx,color}
\usepackage{epic}
\usepackage{rotating}
\usepackage{subfigure}
\usepackage{setspace}
\usepackage{aas_macros}
\usepackage{natbib}
\usepackage{pslatex}

\title[Is acetylene essential for carbon dust formation?]
{Is acetylene essential for carbon dust formation?}
\author[H. Dhanoa \& J.M.C. Rawlings]{H. Dhanoa$^{1}$\thanks{E-mail:
hd@star.ucl.ac.uk(HD); jcr@star.ucl.ac.uk(JCR)}and J.M.C. Rawlings$^{2}$\\
Department of Physics and Astronomy, University College London,
Gower Street, London, WC1E 6BT}

\begin{document}


\pagerange{\pageref{firstpage}--\pageref{lastpage}} \pubyear{2013}

\maketitle

\label{firstpage}

\begin{abstract}
We have carried out an investigation of the chemical evolution
of gas in different carbon-rich circumstellar environments. Previous studies
have tended to invoke terrestrial flame chemistries, based on acetylene
(C$_2$H$_2$) combustion to model the formation of carbon dust, via
Polycyclic Aromatic Hydrocarbons (PAHs).
In this work we pay careful attention to the accurate calculation of the
molecular photoreaction rate coefficients to ascertain whether there is a
universal formation mechanism for carbon dust in strongly irradiated
astrophysical environments.
A large number of possible chemical channels may exist for the formation of
PAHs, so we have concentrated on the viability of the formation of the
smallest building block species, C$_2$H$_2$, in a variety of carbon-rich
stellar outflows. C$_2$H$_2$ is very sensitive to dissociation by UV
radiation. This sensitivity is tested, using models of the time-dependent
chemistry.
We find that C$_2$H$_2$ formation is sensitive to some of the physical
parameters and that in some known sources of dust-formation it can never
attain appreciable abundances. Therefore multiple (and currently
ill-defined) dust-formation channels must exist.
\end{abstract}


\section{Introduction}

The observational properties of dust and the physical consequences of its
presence in the interstellar medium and extremely well-known and
well-documented \citep{D03}.
However its composition, structure and size-distribution are still subjects
of much discussion.
Even though there are no direct samples of interstellar dust, carbonaceous
dust deposits found in meteorites give us an indication of what structure and
composition we might expect. These grains consist of a core of
microcrystalline aromatic moieties surrounded by a crystalline
graphite mantle \citep{T97}.
Significantly, the precise chemical kinetic pathways leading to the
(relatively simple) case of carbon-dust formation are almost entirely
speculative.

The recent observations of higher than expected dust masses around high
redshift quasars (\citealt{DGJ07}; \citealt{BCC03}) and local supernovae
\citep{M11}, has caused a resurgence of interest in the issues
of dust formation mechansims and efficiencies.
In the local universe, dust is observed mainly in two
environments: dense interstellar clouds and circumstellar media. The
main sources of interstellar dust are believed to be evolved red giants,
novae and supernovae (\citealt{D04}; \citealt{S77}; see \citealt{T08} for a
review).
There are thus two questions that we address in this paper: (i) what is the
chemical pathway for the formation of macromolecular species that leads to the
formation of dust nucleation sites?, and (ii) what range of physical
parameters determine the viability of these pathways?

For the sake of (relative) simplicity we limit our discussion to the formation
of carbon dust in carbon-rich circumstellar environments, making the
simplifying assumption that the oxygen is completely locked up in the
relatively inert form of CO. In such an environment the C$:$O ratio - 1 is
thus a crucial parameter.
However, it should be noted (see below) that this usually-assumed approximation
has been challenged by relatively recent observational and modelling 
efforts.

Some of the earliest attempts at modelling carbon dust formation concentrated
on novae, some of which are seen to produce very optically thick (carbon) dust
shells on timescales of days. \citet{R88}
developed a proto-PDR model of the chemistry in novae ejectae,
and found that molecule formation requires the presence of H$_2$ and carbon
neutrality (CI). The presence of C$^+$ (and the carbon ionization continuum)
inhibits molecule formation. Chemical equilibrium is established very fast
(within $\sim 10$s). \citet{RW89}, hereafter RW89,
extended this and proposed a non-LTE chemical kinetic model of (carbon)
dust formation in novae ejectae which proceeds via a limited hydrocarbon
chemistry - totally dominated by the photophysics, and requiring all oxygen to
be locked up in CO. This latter finding has been challenged - both by the
observations that suggest that carbon dust is formed in oxygen rich (O$>$C)
environments \citep[e.g.][]{ADS94} - and also models of the chemistry in nova
outflows (\citealt{PR04}).
In these models C$_2$ (for which the more significant of the two
pre-dissociation bands is shielded against photodissociation in a CI region)
is the key building block species, rather than C$_2$H$_2$ which is
susceptible to photodissociation.
As such, nucleation of dust was found to be possible only in a CI region, but
requires some source of ionization to drive the chemistry.

\citet{PR04} extended the work by RW89 by including
oxygen and nitrogen chemistry and updated the photoreaction rates.
Surprisingly, they found that the overall chemistry is not as dependent on the
photochemistry as earlier studies had suggested and that carbon dust could be
formed even when all of the oxygen is not `locked' in CO.
The work highlighted that CN, CO and C$_2$ are key molecular species in the
formation pathways of larger species.
Although C$_2$H$_2$ was included its chemistry was not closely scrutinised.

More generally, it has been proposed that Polycyclic Aromatic Hydrocarbon
molecules (PAHs) are key intermediates in carbon dust formation and act as
nucleation sites for dust grains (\citealt{H96};
\citealt{CBT92}). This inferred link is due to the the
close relationship between the molecular structure of PAHs and the
carbon structure of graphite. Also, observations of the far-IR
continuum due to carbon dust are accompanied by strong PAH features
\citep{T08}.
However, some caution should be exercised with the interpretation of these
features. For example,
PAH (and/or hydrogenated amorphous carbon) emission is also detected in some
carbon-dust producing novae but in all cases it is only seen well {\it after}
dust formation and in an environment when it would be susceptible to
photodissociation.
This indicates that - for novae at least - the PAHs are transitory products
of dust destruction, rather than intermediates to dust formation \citep{ER08}.

\citet{GS87} proposed chemical mechanisms
required to form PAH molecules in stellar winds. The acetylene
molecule (C$_2$H$_2$) was found to be crucial in the formation pathway.
An important contribution came from
\citet{FF89}, hereafter (FF89).  They investigated a
mechanism of PAH formation, using a chemical kinetic approach. The
chemistry network was based on soot production in terrestrial
hydrocarbon flames, which they applied to astrophysical conditions,
i.e. much lower pressure and densities. They confirmed that the
complex network of chemical reactions can essentially be described by a
sequence of hydrogen abstraction followed by acetylene addition.

Acetylene does not possess a permanent dipole moment and so does not have a
rotational spectrum observable at radio wavelemgths. It is therefore somewhat
elusive; - detections rely on rovibrational transitions, observable in the
mid-infrared. For this reason, detections tend to be limited to the warm gas
($\sim 100-1000$K) along the lines of sight towards young stellar objects
(e.g. \citealt{BDL03}).
There is also strong evidence for the presence of C$_2$H$_2$ in evolved carbon
stars; the most well known of which - IRC+10216 - exhibits a forest of
C$_2$H$_2$ lines in the 11-14$\mu$m window (\citealt{F08}).
Indeed, after H$_2$ and CO, C$_2$H$_2$ is determined to be the most abundant
species in the gas - with a fractional abundance of $\sim 8\times 10^{-5}$ -
although this is primarily detected {\it outside} the dust-formation zone and
so is not necessarily associated with the dust-formation process.
There may even be some evidence of C$_2$H$_2$ freeze-out into icy mantles at
larger radii in IRC+10216, and an icy mantle-based origin for interstellar
C$_2$H$_2$ is strongly suggested by Spitzer observations of the correlation
between C$_2$H$_2$ and gas-phase CO$_2$ towards Cepheus A East
(\citealt{SGN07}).

\citet{CBT92} addressed the issue of PAH chemistry and dust
formation in the carbon-rich envelopes of late AGB stars; driven by stellar
pulstations and (strong, density enhancing) shocks.
Although extremely dense, the gas is much cooler and not subject to the
intense FUV radiation field that is present in a nova, so the chemical model
was based on a terrestrial neutral gas acetylenic-burning soot chemistry.
In such environments the carbon budget is largely split between CO and
C$_2$H$_2$.
The chemistry was largely based on well-studied flame chemistries (eg. FF89),
expanded and augmented to include free radicals.
It was found that C$_2$H$_2$ is the key species in macro-molecular growth and
(especially) ring closure/cyclization to form aromatic hydrocarbon rings, as
well as the polymerization to multiple ring PAHs.
The formation of the first aromatic ring is often recognised as the main
`bottleneck' in macro-molecular growth/dust nucleation.
In this, as in later studies \citep[e.g.][]{ChC99}, ring closure was found 
to proceed
following reactions of two propargyl radicals (C$_3$H$_3$) or via a lesser
channel involving reaction of acetylene with 1,3 butadiyne (C$_4$H$_2$)
(e.g. see Figs. 3 \& 4 in that paper). In either case C$_2$H$_2$ initiates the
chemistry.
Reaction networks need to include both two-body and three-body reactions, but
did not include photoreactions, or an ion-molecule chemistry.
Although they were able to identify the presence of a temperature window
(900-1100K) in which the formation of PAHs is possible, the formation
efficiency is too low - possibly due to the lack of inclusion of the effects
of a local UV radiation field and/or an ion-molecule chemistry.

Considering a different class of object \citet{W02}
modelled the formation of benzene (C$_6$H$_6$) in a
protoplanetary nebula (CRL618) where the chemistry is characterised by high
densities, temperatures and ionization rates - although photoreactions were
significantly inhibited throughout the molecule-formation zone due to the
assumed presence of dust. This chemistry is somewhat different to the
Cherchneff et al. schemes due to the presence of a degree of ionization and
an efficient ion-molecule chemistry. However, even in these conditions,
C$_2$H$_2$ plays a crucial role - mainly through the initiating reaction of
\[ {\rm HCO^+ + C_2H_2 \to products...} \]
All of the above models and reaction schemes have made extremely simplistic
assumptions about the radiation field and the photoreaction rates; either
by only considering the chemistry in low luminosity environments (or external
to an optically thick dust-formation zone) - in which case they are ignored, or
by simply `scaling up' interstellar photoreaction rates in a way that does not
properly take into account the specific nature of the radiation field and/or
the significance of ionization continua.
The overall aim of this work is therefore to include an ion-molecule
chemistry and a careful re-calculation of the photoreaction rates using local
radiation fields and accurate cross-sectional data
to re-assess the viability of C$_2$H$_2$ formation in a variety of
carbon-rich circumstellar environments.
As C$_2$H$_2$ is fundamental to the formation of PAH molecules, we
try to determine if the amount of
C$_2$H$_2$ produced in carbon rich AGB and nova environments is
enough to seed dust formation. We use a specially adapted chemical
network for high temperatures and densities, paying particular
attention to a more accurate account of the photochemistry involved
in non-interstellar conditions, which (as explained above) previously had 
been highly over-simplified.
We investigate a broad range of dust-forming objects which includes carbon
stars with effective temperatures higher than AGB stars (e.g. CH subgiants),
novae and R Coronae Borealis stars.
By doing this we aim to establish the range of parameter space
within which effective C$_2$H$_2$ formation (and by inference PAH
formation) is possible.

In section \S\ref{sec:model}, we describe the model that was implemented.
Section \S\ref{sec:parameters} specifies the physical and chemical parameters
used for the three different carbon rich circumstellar environments that
we have modelled. We describe the results and analysis in section
\S\ref{sec:results}. Finally, in section \S\ref{sec:conc} we summarise our
results and their implications.

\section{The Model}
\label{sec:model}

The nucleation sites for dust grain formation are formed in the gas
phase in a cooling, expanding atmosphere, such as in the outflow
from a red giant \citep{CWF94}. Therefore if PAHs are
the key intermediate molecules to carbonaceous dust they should also
be formed in the stellar winds that feed the circumstellar
environment. We therefore model the time-dependent chemistry that
occurs in an appropriately located gas parcel. This will be the part
of the envelope whose temperature lies within the range that is appropriate
for dust formation - corresponding to the `condensation radius' in
previous studies (eg. \citealt{CW76}).

As explained above, the main chemical pathway to form a PAH molecule starts
with the polymerization of C$_2$H$_2$ \citep{GS87} and we
therefore identify acetylene as being
the key molecule in the formation of larger species, such as PAHs.
We investigate the viability of acetylene formation as being the controlling
factor in the formation of dust nucleation sites via PAHs as intermediate
macro-molecules.


The gas in stellar outflows is dense (n$>10^9$cm$^{-3}$) and warm
(T$>1000$K), so that (providing
the gas is not {\em too} hot, such that bond-breaking, collisional
dissociation reactions can occur), molecules can be formed
efficiently and quickly. As the gas expands and its density falls,
the chemical reaction rates are quenched.
However, the geometrical dilution timescale is typically many orders of
magnitude larger than the chemical kinetic timescale
(\citealt{R88}) and this is not a critical factor in
determining the dust-formation efficiency. Despite the possible presence of
shocks etc. - and although the gas is very far from thermal equilibrium,
the characteristic timescale of the outflow is much larger than the
chemical kinetic timescale. We therefore assume that both the
circumstellar environment and the stellar wind are spherically
symmetric around the star and that the physical conditions (density,
temperature, radiation dilution factor etc.) are not time dependent
and remain constant.

The chemistry is evolved to chemical equilibrium which,
as argued above, occurs on a timescale that is very much shorter than that
for which changes in the physical conditions occur.
In practice, equilibrium was determined from the requirement that the
abundances of all species changed by $<1\%$ on any time step and typically
was achieved on timescales of seconds.
We have investigated the sensitivity of the equilibrium abundances to the
physical conditions. The key free parameters that we have
varied are: the total abundances of the elements, the photospheric
temperature of the star, the density and the temperature of the gas.
Other parameters, such as the dilution factor for the radiation field, are
taken to be defined by the observed location of dust formation.
The elemental abundances are given in Table~4.
The cosmic ray ionization rate is set to $10^{-17}$s$^{-1}$, which is
representative of the assumed galactic background rate
\citep{D06}.

\subsection{The Chemistry}

The chemical network includes 1537 reactions between 102 gas-phase
species involving 8 elements (H, He, C, N, O, S, Na and Si). These
are listed in Table~1.

\setcounter{table}{0}
\begin{table}
\caption{Chemical Species}
\begin{tabular}{|c|}
\hline
H, H$_2$, H$^+$, H$^-$, H$_2^+$, H$_3^+$ \\
C, C$^+$, C$^-$, CO, CO$^+$, CH, CH$^+$, CH$_2$, CH$_2^+$, CH$_3$, CH$_3^+$ \\
CH$_4$, CH$_4^+$, C$_2$, C$_2^+$, C$_2$H, C$_2$H$^+$, C$_2$H$_2$,
C$_2$H$_2^+$\\
N, N$^+$, NH, NH$^+$, NH$_2$, NH$_2^+$, NH$_3$, NH$_3^+$, NH$_4^+$, N$_2$,
N$_2^+$, N$_2$H$^+$ \\
O, O$^+$, O$^-$, O$_2$, O$_2^+$, OH, OH$^-$, OH$^+$, H$_2$O, H$_2$O$^+$,
H$_3$O$^+$ \\
HCO, HCO$^+$, H$_2$CO, H$_2$CO$^+$, CO$_2$, CO$_2^+$ \\
CN, CN$^+$, CN$^-$, HCN, HCN$^+$, HNC, HCNH$^+$ \\
NO, NO$^+$, HNO, HNO$^+$, NO$_2$, NO$_2^+$, OCN \\
He, He$^+$, HeH$^+$, Na, Na$^+$, e$^-$ \\
S, S$^+$, S$^-$, HS, HS$^+$, H$_2$S, H$_2$S$^+$, H$_3$S$^+$ \\
CS, CS$^+$, HCS, HCS$^+$, H$_2$CS$^+$, SO, SO$^+$, SO$_2$, SO$_2^+$, OCS,
OCS$^+$ \\
NS, Si, Si$^+$, SiH, SiH$^+$, SiH$_2^+$, SiO, SiO$^+$, SiOH$^+$, SiO$_2$ \\
\hline
\end{tabular}
\label{tab:species}
\end{table}

The chemical complexity of the reactants was limited as we
only concentrate on the initiating steps in the formation of larger
molecular species.

The ratefile is an adapted set of reactions, applicable
to high densities, high temperatures and intense radiation fields.
Many of the reactions and rate coefficients (k$_i$) were taken from the UMIST
database for Astrochemistry \citep{W07} where the formulation
for the rate coefficients is valid in the operative temperature range.
Additional data was taken from databases that are relevant to high
temperature/high density environments \citep[e.g.][etc.]{RDB93,L02,PR04}.
Although the densities are very much larger than those applicable to
interstellar studies, they are not so high ($\geq 10^{12}$ cm$^{-3}$)
that three-body reactions need to be taken into account.

Reaction types that are included in the chemistry are: charge transfer,
ion-molecule, radiative association, neutral exchanges, radiative
recombination, dissociative recombination and negative ion reactions.
Reactions
which are particularly significant in addition to those applicable to
interstellar clouds are neutral-neutral reactions, reactions with
significant activation energy barriers, and reactions between
ro-vibrationally excited species, due to the high temperatures
that exist in circumstellar outflows.

The main formation and destruction reactions for C$_2$H$_2$ are given in
Tables~2 and 3.

\setcounter{table}{1}
\begin{table}
\caption{Main formation reactions of C$_2$H$_2$, rates from
\citet{W07}}
\begin{tabular}{|c|c|c|}
\hline
&Reaction & Reaction rate\\
\hline
1& CH$_2$ + CH$_2\rightarrow$ C$_2$H$_2$ +2H &
k$_1$=$1.80\times 10^{-10}\mbox{e}^{-400 K/T}$\\
2& CH$_2$ + CH$_2\rightarrow$ C$_2$H$_2$ +H$_2$
 & k$_2=2.63\times 10^{-9}\mbox{e}^{-6013 K/T}$ \\
3&H$^-$ + C$_2$H $\rightarrow$ C$_2$H$_2$ + e$^-$
 & k$_3=1.00\times 10^{-9}$\\
4& C$^-$ + CH$_2 \rightarrow$ C$_2$H$_2$ + e$^-$
 & k$_4=5.00\times 10^{-10}$\\
5& C + CH$_3 \rightarrow$ C$_2$H$_2$ + e$^-$
 & k$_5=1.00\times 10^{-10}$\\
 \hline
\end{tabular}
\label{tab:form}
\end{table}

\setcounter{table}{2}
\begin{table*}
\centering
 \begin{minipage}{140mm}
  \caption{Main destruction reactions of C$_2$H$_2$, rates from
\citet{W07}}
\begin{tabular}{|c|c|c|}
\hline
&Reaction & Reaction rate\\
\hline
6& C + C$_2$H$_2 \rightarrow$ C$_3$H + H & k$_{6}=1.45\times
10^{-10}\left(\frac{T}{300}\right)^{-0.12}$\\
 7&C + C$_2$H$_2 \rightarrow$ C$_3$ + H$_2$
& k$_{7}=1.45\times 10^{-10}\left(\frac{T}{300}\right)^{-0.12}$\\
8&H$^+$  C$_2$H$_2 \rightarrow$ C$_2$H$^{+}_2$ + H
& k$_{8}=5.40\times 10^{-10}$\\
 9&H + C$_2$H$_2 \rightarrow$ C$_2$H +H$_2
$& k$_{9}=3.80\times 10^{-10} \mbox{e}^{-\frac{13634}{T}}$\\
10& O + C$_2$H$_2 \rightarrow$ CO + CH$_2$ & k$_{10}=8.39\times
10^{-12}\left(\frac{T}{300}\right)^{1.03}
\mbox{e}^{-\frac{1197}{T}}$\\
11& O + C$_2$H$_2 \rightarrow$ C$_2$H +OH
& k$_{11}=5.30\times 10^{-9} \mbox{e}^{-\frac{8520}{T}}$\\
12& He$^+$ +C$_2$H$_2 \rightarrow$ CH$^+$ + CH +He &
k$_{12}=7.70\times10^{-10}$\\
13& He$^+$ +C$_2$H$_2 \rightarrow $C$^+_2$ + He + H$_2$
& k$_{13}=1.61\times 10^{-9}$\\
14& He$^+$ +C$_2$H$_2 \rightarrow$ C$_2$H$^+$ + He + H
& k$_{14}=8.75\times 10^{-10}$\\
15&He$^+$ +C$_2$H$_2 \rightarrow$ C$_2$H$_2^+$ + He
& k$_{15}=2.54\times 10^{-10}$\\
16&C$^+$ + C$_2$H$_2 \rightarrow$ C$_3$H$^+$ + H& k$_{16}=2.20\times 10^{-9}$\\
 17&S$^+$ + C$_2$H$_2 \rightarrow$ HC$_2$S$^+$ H
& k$_{17}=9.50\times 10^{-10}$\\
18& Si$^+$ + C$_2$H$_2 \rightarrow$ SiC$_2$H$^+$ + H &
k$_{18}=1.30\times 10^{-10}\left(\frac{T}{300}\right)^{-0.71}
\mbox{e}^{-\frac{29}{T}}$\\
19&Si$^+$ + C$_2$H$_2 \rightarrow$ SiC$_2$H$_{2}^+$ + photon &
k$_{19}=2.50\times
10^{-10}$\\
20&Si + C$_2$H$_2 \rightarrow$ SIC$_2$ +H$_2$ & k$_{20}=1.00\times
10^{-13}\left(\frac{T}{300}\right)^{-1.00}$\\
 \hline
\end{tabular}
 \end{minipage}
\label{tab:dest}
\end{table*}

\subsubsection{Photoreactions}

In circumstellar environments, irradiated by the central star, the
UV radiation field is typically very much stronger and has quite a
different spectral shape to that whch pertains in the interstellar
medium. It is therefore important that we pay particular attention
to the correct evaluation of the photo-reaction rate coefficients:
The rates for all photodissociation, photoionisation and
photodetachment reactions are calculated {\em ab initio}.

In these conditions, the key factor in determining the
viability of any chemical kinetic pathway that leads to
dust-formation is therefore the stability of the intermediates against
photodissociation/photoionisation by the stellar radiation field.

The calculation of the photorates by a time-dependent non-interstellar
radiation field in an expanding atmosphere is a non-trivial task.
There are two components to the calculation of the photoreaction
rate-coefficients in the local radiation field:
\begin{enumerate}
\item a basic `photospheric' rate, calculated for the specific photoreaction
and radiation field
\item a time-dependent geometrical dilution, due to the expansion of
the wind
\end{enumerate}

In earlier models, and for the sake of simplicity, photoreaction
rates were calculated by scaling the unshielded interstellar
photorates by a ratio of the stellar flux to that of the
interstellar radiation field, either by a single scaling factor, or by
a value appropriate to a single characteristic
wavelength for each reaction in question (\citealt{R88}).
In this study we use a more accurate approach that can encompass the variety
of situations that we wish to investigate: Photoreactions are
calculated by fully integrating the known cross-sections and
oscillator strengths (for line transitions) over the specified
radiation field. Following \citet{VD88} we treat the contributions
to the photodissociation rate arising from continuous absorption and those
arising from absorption in discrete transitions as separable. These can be
written as
\begin{equation}
k_{pd}^{cont}=\int\sigma(\lambda)I(\lambda) \, d\lambda \,
\mbox{s}^{-1}
\end{equation}
and
\begin{equation}
k_{pd}^{line}=\frac{\pi
e^{2}}{mc^{2}}\lambda_{ul}^{2}f_{ul}\eta_{u}x_{l} I(\lambda_{ul}) \,
\mbox{s}^{-1}
\end{equation}
respectively, where $k$ is the photoreaction rate coefficient, $\lambda$
is the wavelength, $I_\lambda$ is the flux, $\sigma(\lambda)$ is the
cross-section for the reaction, $f_{ul}$ is the oscillator strength and
the other symbols have their usual meanings. Photoreaction
cross-sectional data and oscillator strengths were taken from a variety of 
sources \citep[e.g.][]{VD88,RDB93} as described in \citet{RR12}.
We assume that the gas is within the carbon-neutral zone
(equivalent to being behind the photodissociation region: As was
shown in \citet{R88}, complex chemistries are unsustainable
in C{\sc ii} regions). We therefore assume that H$_2$ and CO are fully
self-shielding and that carbon photoionization is effectively
suppressed.
The lower limit of the integral in equation 1 is therefore set to the 
ionization threshold for C{\sc i} (1102\AA).

This study is too generalised to consider the radiation fields of
individual sources so we instead opt to characterise the radiation
field by that of an (appropriately diluted) single temperature
black-body ($T_{BB}$). In addition, as we have investigated a wide
range of $T_{BB}$ which may also be time-dependent, we have utilised
a parametric fit to the dependence of the photoreaction rates to
$T_{BB}$. Thus, after calculating the photospheric rates for several
values of $T_{BB}$ we fit a simple quadratic to the logarithm of the
rates:
\begin{equation}
\log(k) = a + b.\log T_{BB} + c.\log (T_{BB})^2
\end{equation}
where the coefficients $a,b$ and $c$ are specific to each reaction
and are defined for a certain temperature range. The ranges used are
as specified in Table~4.

For the geometrical dilution we assume simple spherical symmetry,
in which case, at a distance $R$ from the star, the the dilution factor is
given by
\begin{equation}
W=\left(\frac{R_\star}{2R}\right)^2\mbox{     }R\gg
R_\star,\label{eqn:dil}
\end{equation}
where $R_\star$ is the photospheric radius of the star.

To ascertain the viability of the efficient formation of C$_2$H$_2$, we
will need to study
the chemistry in the vicinity of the `condensation radius' where
dust grains start to form. In the case of AGB envelopes, this is
also where outflow acceleration due to radiation pressure on newly
formed dust grains starts. For AGB winds, the radius of
condensation, $R_c$, is given by \citet{CBT92};
\begin{equation}
R_c=R_{\star}\left(1-\left[1-2\left(\frac{T_{gas}}{T_{phot}}\right)^{4.85}
\right]^{2}\right)^{-0.5}.\label{eqn:rc}
\end{equation}
Substituting this into equation (\ref{eqn:dil}), the dilution factor
becomes
\begin{equation}
\mbox{W}=\frac{1}{4}\left\{1.0-\left[1.0-2.0\left(\frac{T_{gas}}{T_{phot}}
\right)^{4.85}\right]^2\right\}.\label{abgW}
\end{equation}
which is only dependent on the ratio of the temperature of the
photosphere ($T_{phot}$) to that in the gas ($T_{gas}$).
In the case of the AGB winds, the temperature of the gas parcel is set at 
1000K, which is the midpoint of the temperature window for PAH formation
(900-1100 K) found by \citet{CBT92}.

Finally, we include a (small) contribution to the photorate deriving
from the local internally generated UV field induced by cosmic ray
ionizations \citep{PT83,G89}.

\section{Physical and Chemical Parameters}
\label{sec:parameters}

We have investigated four different carbon-rich (C$>$O) astrophysical
environments. We have studied acetylene formation in the outflows from:
\begin{enumerate}
\item Thermally pulsing asymptotic giant branch (AGB) stars,
\item Novae,
\item G/K-type Carbon stars, and
\item R Coronae Borealis-type stars.
\end{enumerate}
The physical conditions are quite different in these four classes of objects,
yet they are each known to be prolific sources of carbon dust.

Firstly, we have extended the work of \citet{CBT92}, which modelled PAH 
formation in the post-shock gas in thermally pulsing asymptotic
giant branch (AGB) stars. These are well known dust production sites.
Including a more complete photochemistry has allowed us to determine
more accurately the stability of C$_2$H$_2$ in this environment.
We have applied our model to examine the viability of PAH formation
in the ejecta around novae, some of which (but not all) are also known 
to be efficient dust-producers. 
We have also investigated the outflows from stars that are somewhat
hotter than AGB stars, so that we can ascertain the possibility of soot
chemsitries operating in these circumstellar environments. 
Finally, we have modelled the conditions required for acetylene-driven dust
formation in R Coronae Borealis (R Cor Bor) stars, which are well-known to 
be prolific dust-producers. 
The initial conditions selected for each of these four scenarios are 
discussed below.

There are considerable variations (and uncertainties) in the 
observationally inferred elemental abundances for each source type.
However, the C$:$O ratio is a key parameter. This determines the efficiency 
of the hydrocarbon chemistry and, hence (whatever reaction pathway is 
followed) the viability of carbon dust formation. 
To effect the requisite variotions in the C$:$O ratio we have adopted cosmic
abundance values for all elements other than carbon;
0.85, $6.0\times 10^{-5}$, $4.6\times 10^{-4}$, $1.4\times 10^{-5}$,
$3.5\times 10^{-5}$ and $3.2\times 10^{-5}$ for He, N, O, S, Na and Si
respectively. The abundance of carbon is then varied to give the (observed) 
C$:$O ratio as specified in Table~4.

\setcounter{table}{3}
\begin{table*}
 \centering
 \begin{minipage}{140mm}
  \caption{Physical and chemical parameter ranges for the four different
  cases being investigated.}
  \begin{tabular}{@{}lcccc@{}}
  \hline
      & AGB & Nova & Stars of spectral Class G-K& R Cor Bor stars \\
 \hline
 [C/O] & 1.5 & 21 & 1.5 & 7.2 \\
 Density Range (cm$^{-3}$) & 10$^9$--10$^{11}$ & 10$^9$--10$^{11}$ &
 10$^9$--10$^{11}$ & 10$^9$--10$^{12}$  \\
 Photosphere Temperature (K) & 1500--3500  & 15,000--25,000  & 3750--6000
 & 4500-8000 \\
 Gas Temperature (K) & 1000  & 2000  & 1000 & 1000 \\
 Dilution Factor & see eqn (\ref{abgW})& 5.59x10$^{-5}$ & see eqn (\ref{abgW}) & 6.25x10$^{-4}$ \\
 \hline
\end{tabular}
\end{minipage}
\label{tab:param}
\end{table*}

\subsection{Case \sc i: AGB Stars}

High carbon-oxygen ratios ($>2$) are unlikely in the Galaxy (although
common in the Large Magellanic Cloud) and are
excluded. Lower ratios of $C:O\sim 1.5$ are expected for most carbon
stars \citep{BKR01}.
\citet{L86} found slightly subsolar values for the oxygen and
nitrogen abundances in AGB stars, but our approximation of using solar
abundances is essentially valid. 
This is consistent with the abundances used by \citet{CBT92} to model an 
AGB star that is similar to the well-known source IRC+10216. 
We can therefore compare our results to \citet{CBT92} and distinguish 
the effects of including a complete photochemistry. In a study of of 390 
carbon rich stars by \citet{BKR01}, the highest effective temperature
evaluated was 3870 $\pm$ 1080 K. We have therefore investigated 
photospheric temperatures in the range T=1500-3500K. The initial conditions 
and parameter ranges that we consequently adopt for AGB outflows are given 
in Table~4. 

\subsection{Case \sc ii: Novae}

\setcounter{table}{4}
\begin{table*}
 \centering
 \begin{minipage}{140mm}
\caption{Table of nova parameters}
\begin{tabular}{|p{6cm}|c|}
\hline
Nova Parameter & 30 days post-outburst\\
\hline
Rate of magnitude decline (optical), $\dot{m}_{\mbox{v}}$ & 0.03 mag d$^{-1}$\\
Luminosity  of nova, L$_{\mbox{nova}}$ & 31080 \\
Photospheric radius, r$_{\mbox{phot}}$ & 25 R$_\odot$ (=R$_{p,0}$)\\
Photospheric temperature, T$_{\mbox{phot}}$& 20,100 K\\
\hline
\end{tabular}
\end{minipage}
\label{tab:nova}
\end{table*}

It has long been established that dust is sometimes formed, rapidly and
efficiently, in nova outflows \citep[e.g.][]{G77}. In the most extreme
circumstances, a steep decrease in the visual emission occurs, accompanied by 
a strong rise in thermal infrared emission - as an optically thick dust shell 
is formed \citep{CW76}. 
Following \citet{R88}, we adopt the following relations between the physical 
parameters:
\begin{align}
& \Delta m_{\mbox{v}} =\dot{m_{\mbox{v}}}\times t_0, \\
& \mbox{L}_{\mbox{nova}}=1.036\times 10^6 \times{\dot{m_{\mbox{v}}}},\\
& \mbox{r}_{\mbox{phot}}\sim 9.97\times10^9\times \mbox{L}_{\mbox{nova}},\\
& \mbox{T}_{\mbox{phot}}=15280\times10^{(\frac{\Delta
m_{\mbox{v}}}{7.5})},
\end{align}
where $\dot{m_{\mbox{v}}}$ is the rate of magnitude decline in the
optical, $\mbox{r}_{\mbox{phot}}$ is the initial photospheric
radius, $\mbox{L}_{\mbox{nova}}$ is the nova luminosity and $t_0$ is
the number of days post outburst.\smallskip

The values for the rate of visual decline from maximum ($\dot{m_v}
=0.03$ mag d$^{-1}$) and time at start of calculation ($t_0$= 30
days post outburst) were adopted following
\citet{RW89}. Novae can be classed by the time taken
for a nova to diminish by 2 magnitudes below maximum visual
brightness. Generally, `slower' novae (i.e. those with smaller values of 
$\dot{m_v}$) are more effective at producing optically thick shells of 
carbon dust. Parameters for a typical dust-producing nova are given in 
Table~5 and are used in our model.
The parameter range investigated and initial conditions are given in 
the Table~4. 
The fractional abundance of carbon is much higher than in the case of 
AGB stars \citep{RW89}. This is due to the thermonuclear processing of the gas 
and dredge-up of white dwarf material that occurs following the thermonuclear
runaway that drives the nova outburst.
The upper limit of density was inferred from observations \citep[e.g.][]{R88}.

\subsection{Case \sc iii: Carbon Stars in Spectral Class G-K}

We investigate stars with photospheric temperatures in
the range 3750--6000K. Low mass main sequence stars are not likely
to have a high C$:$O ratio.
However, CH subgiants are low luminosity peculiar
giants, which have a high C$:$O ratio in their atmosphere as a
result of mass transfer from an evolved companion
\citep{SCL93}. The upper limit of C$:$O observed in
these stars is 1.5 \citep{LB82}. The parameter ranges chosen to represent 
these objects are given in Table~4. 
We have optimised the parameters for dust production by only considering the
highest values possible for the C$:$O ratio and the gas density. 

\subsection{Case \sc iv: R Coronae Borealis stars}

R Coronae Borealis stars (RCB's) are yellow supergiants which are hydrogen deficient and carbon rich. These stars show irregular declines in visual brightness caused by the production of  thick dust clouds \citep{Cr07}. 
There are also variations in observed colour that correspond to the minimum in
visual brightness and which are attributed to pulsations \citep{Ef88}. 
These stars have fast stellar winds that feed into a thin dust shell found 
hundreds of stellar radii away \citep{W86,F86}. 
The optical properties of the ejected dust differs from the ISM, with a
UV absorption peak at 2500 \AA\, \citep{J12}. Models indicate that this
corresponds to small glassy or amorphous carbon grains (soot) formed in a hydrogen-poor environment \citep{H84}.\smallskip

The dust formation mechanism is not at all well understood and there has been
much debate as to where the dust condensation occurs. 
There are two main models: i) the first assumes the grains are formed at 
$\sim$20 R$_{\star}$, and ii) the second proposes that dust is formed in the
photosphere of the star and moves quickly away due to radiation pressure 
\citep{PG63}. 
The rationale behind the first method is that the stellar surface temperature
greatly exceeds the grain condensation temperature for carbon \citep{F86,F88}.
\citet{Wo96} proposed that the pulsations stronger than a certain critical
amplitude could cause atmospheric shocks to form in a process similar to that proposed for AGB outflows. This yields post-shock gas temperatures and densities
of T$\sim$1000-1500~K and n$\sim  10^8 -10^9$ cm$^{-3}$, which may be propitious
for dust formation. In fact, observations by \citet{J12} indicate that there 
are two populations of grains which may indicate that both formation processes are occurring.
\smallskip

We investigate C$_2$H$_2$ formation on the assumption of the first model, i.e.
molecule/dust formation at a radius of 20 R${_\star}$ from the star. 
The temperature of the gas in this region is observed to be $<$ 1500 K \citep{GS92}. The effective temperature of RCB's range from 4,000-20,000 K, with mean
values of 6000-8000 K for dust-forming sources \citep{LR94}. 
The parameter ranges adopted for this scenario are given in 
Table~4. The abundances for carbon, nitrogen and helium were obtained from 
\citet{GH09}, whereas all other abundances are assumed to be solar.

\section{Results}
\label{sec:results}

C$_2$H$_2$ formation is a necessary, but not sufficient, criterion for
carbon dust formation via PAH chemistry. Therefore we need to establish the
criterion for a positive result, i.e. the minimum fractional abundance of
C$_2$H$_2$ that will explain the requisite abundance of dust
nucleation sites. 
To do this, we follow \citet{RW89}; where the number of grains is equated 
to the number of nucleation sites. Considering the case of a nova, the number 
of grains, N$_g$, inferred from infrared observations is 
$\sim7\times 10^{38}$ \citep{CW76}. The dust shell condenses at a radius 
$r_c\sim8\times10^{14}$ cm, and has a typical thickness of $\sim r_c$/10. 
The shell volume is therefore $V\sim 6\times 10^{44}\,\rm{cm}^3$, implying
a number density of grains of $\sim 10^{-6}$ $\rm{cm}^{-3}$. Note that this 
is a lower limit, subject to the ejecta configuration. The gas density is 
$\sim 10^9$ $\rm{cm}^{-3}$, hence the minimum fractional abundance of 
nucleation sites needed is $10^{-15}$. On the assumption that each C$_2$H$_2$
molecule is efficiently converted into a nucleation site (i.e. that there are 
no kinetic bottlenecks in the formation of larger molecules), we can very
crudely adopt this value as the threshold for the minimum fractional 
abundance of acetylene which can promote a PAH-driven dust formation pathway.
\smallskip

Results from our models are shown in Figures~\ref{fig:agb}-\ref{fig:rcorborstar} as contour plots of the equilibrium abundance of C$_2$H$_2$ as a function 
of density and photspheric temperature.
We discuss the results for the four cases in sections \S \ref{sec:resagb} - 
\S\ref{sec:resrcor}.

\subsection{Case \sc i: AGB}
\label{sec:resagb}

The results for AGB winds are presented in Figure \ref{fig:agb}.
The figure shows that the C$_2$H$_2$ abundance is not strongly sensitive
to the density. There is, however, a strong dependence on the photospheric
temperature. For photospheric temperatures below $\sim$2500K, C$_2$H$_2$ can 
be produced with the requisite efficiency for any density within the range
explored ($\sim 10^9-10^{11}$cm$^{-3}$).
The lack of senitivity to density implies that {\em if} C$_2$H$_2$ leads to PAH
formation, then the production of dust will not be crtically limited to the 
geemetrical dilution of the wind.

\setcounter{figure}{0}
\begin{figure}
\centering{\scalebox{0.6} {\includegraphics{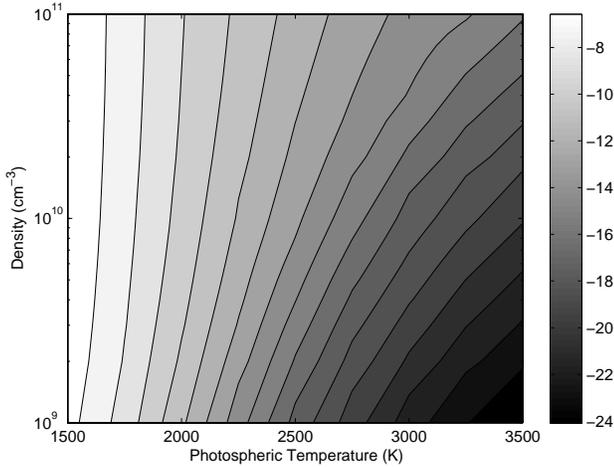}}}
\caption{Contour plot showing the logarithm of the equilibrium
fractional abundance of C$_2$H$_2$ in the circumstellar envelope of
a carbon-rich AGB star. The density and photospheric temperature are
treated as free parameters. All other parameters are fixed and given
values as specified in Table~4.}\label{fig:agb}
\end{figure}

\subsection{Case \sc ii: Nova}

The results for nova outflows are presented in Figure~\ref{fig:nova} 
and show the opposite trends what was found for AGB winds; 
a strong dependence on density, but only a weak dependence on 
photospheric temperature.  
Despite the high abundance of carbon in novae \citep{STS00}, the 
photospheric temperatures are ten times higher than for AGB stars and,
even after significant geometrical dilution is taken into account, 
C$_2$H$_2$ is highly susceptible to photodissociation and never attains 
appreciable abundances. This is a strongly negative result and supports the 
findings of previous studies \citep[e.g.][]{RW89} - C$_2$H$_2$ can not
be produced in irradiated nova environments at abundance levels sufficient 
to seed dust formation via PAH chemistry. 
The only way that a positive result could be obtained is if the ejecta were
extremely dense (n$>10^{13}$cm$^{-3}$) allowing three-body chemical reactions 
to take place, or strongly shielded from the nova radiation field. 

\setcounter{figure}{1}
\begin{figure}
\centering{\scalebox{0.6}{\includegraphics{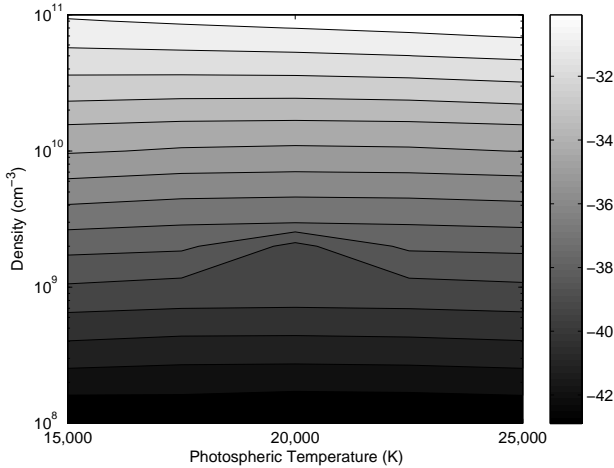}}}
\caption{Contour plot showing the logarithm of the equilibrium
fractional abundance of C$_2$H$_2$ in the ejecta from a nova. The
density and photospheric temperature are treated as free parameters.
All other parameters are fixed and given values as specified in
Table~4.} \label{fig:nova}
\end{figure}

\subsection{Case \sc iii: Stars in Spectral class G-K}

Results for the circumstellar environment of carbon-rich stars with 
photospheric temperatures of 3750--6000K are presented in 
Figure~\ref{fig:gkstar}.
The critical fractional abundance of $\mbox{C}_2\mbox{H}_2$ is not 
reached. However, for the lowest photospheric temperatures (T$<$4500\,K)
and highest densities (n$> 10^{11}$ cm$^{-3}$) the abundance is very much
closer to the threshold than was the case for the nova outflows. We can 
conclude that, whilst the spherically symmetric model produces a negative 
result, a positive result may be obtained in the event of ejecta clumping 
and/or stratification. \smallskip

However, unlike the situation with AGB outflows, this result is not robust 
to density variations so, if the clumps were to dissipate within the 
temperature window for PAH formation then the C$_2$H$_2$ abundance (and 
dust formation efficiency) would be strongly diminished.

\setcounter{figure}{2}
\begin{figure}
\centering{\scalebox{0.6}{\includegraphics[trim= 0mm 0mm 0mm 6mm,
clip]{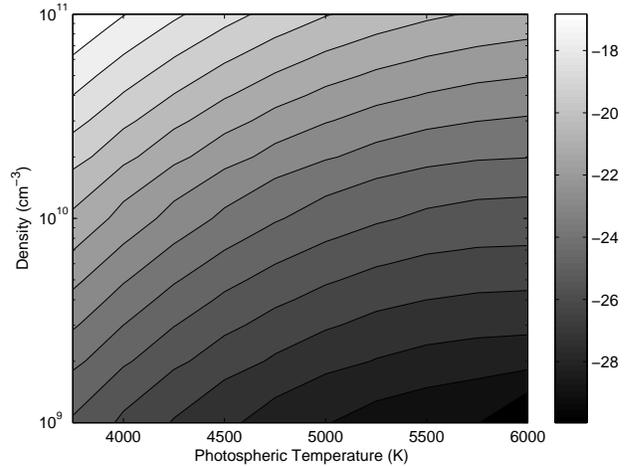}}}
\caption{Contour plot showing the logarithm of the equilibrium
fractional abundance of C$_2$H$_2$ in the circumstellar environment
of a star with spectral class G-K. The density and photospheric
temperature are treated as free parameters. All other parameters are
fixed and given values as specified in Table~4.}
\label{fig:gkstar}
\end{figure}

\subsection{Case \sc iv: R Coronae Borealis stars}
\label{sec:resrcor}

\setcounter{figure}{3}
\begin{figure}
\centering{\scalebox{0.44}{\includegraphics[trim= 10mm 0mm 0mm 6.5mm,
clip]{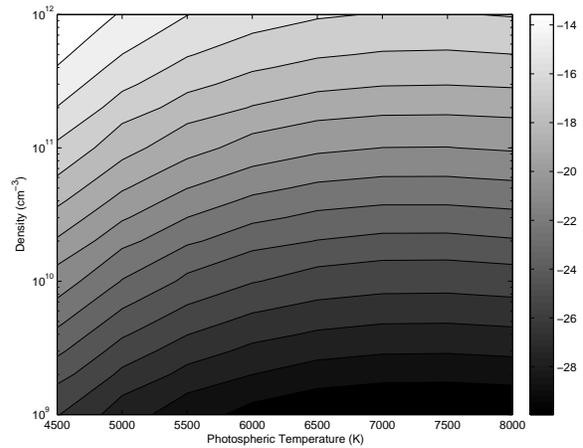}}}
\caption{Contour plot showing the logarithm of the equilibrium
fractional abundance of C$_2$H$_2$ in the circumstellar environment
of a R Coronae Borealis star. The density and photospheric
temperature are treated as free parameters. All other parameters are
fixed and given values as specified in Table~4.}
\label{fig:rcorborstar}
\end{figure}

The results for the gas in a stellar wind located at a radius of 
20 R$_{\star}$ from an R Coronae Borealis star are presented in 
Figure \ref{fig:rcorborstar}. 

A positive result is obtained for part of the range of the physical 
conditions that we have investigated.
The dependencies are similar to those found for the outflows from G-K stars
in that there is a strong dependence on density and a weaker dependence on 
photospheric temperature.
The requisite fractional abundance of C$_2$H$_2$ is obtained for densities 
that $\geq 1-5\times10^{11}$ cm$^{-3}$. For photospheric temperatures 
T$_{phot}< $ 6000\,K, a positive reult can be obtained at lower densities.
We again note that any density enhancements that may exist within the 
outflow will strongly strengthen this result.

\section{Conclusions}
\label{sec:conc}

We have used a simple model of the steady-state chemistry of several different
astrophysical dust forming environments to consider the stability of acetylene 
(C$_2$H$_2$). 
In previous studies of dust formation, based on terrestrial flame chemistries, 
this molecule is a precursor to PAH formation and so its presence at 
appreciable abundance levels can be used to estabish the viability of PAH 
formation and, by inference, dust formation based on PAH condensation chemistry.
However, C$_2$H$_2$ is highly suceptible to photodissoocation by stellar 
ultraviolet.
The novel approach in this study is that we have carefully calculated the 
photodissociation rates for the molecular species involved in the C$_2$H$_2$
chemistry, subject to the local radiation field. Previous studies had made 
major simplifying assumptions in this respect.\smallskip

In each case we have investigated the dependence of the C$_2$H$_2$ abundance as 
a function of gas density and photospheric temperature.
In general, we find that the inclusion of accurate photorates significantly 
restricts the parameter space within which efficient C$_2$H$_2$ formation 
(and hence, presumably, PAH formation) can occur. \smallskip
 
In the case of the winds driven by thermally pulsing AGB stars, we find that
- as in previous studies - C$_2$H$_2$ can be formed very efficiently indeed
and that a PAH chemistry is therefore at least a viable pathway for carbon 
dust formation in these environments. 
This is an interesting results as some previous studies (which have not 
included photochemistry in their modelling) have not been so successful at
producing C$_2$H$_2$ \citep[e.g.][]{CBT92,H96}.
However, we find that this positive result only holds for photospheric 
temperatures that are $<$2500\,K.
There are dust-producing carbon AGB stars with photospheric temperatures 
$\sim 3500$K \citep[e.g.][]{EE95}. If we were to include a more accurate 
physical model of the outflows, including shocks and clumping, 
a PAH chemistry may also be applicable in these objects. Alternatively, other
dust formation mechanisms may be operating.\smallskip 

As compared to the dust-forming environment of AGB stars,
the ejectae of novae are subject to a much stronger radiation 
field with a significantly higher black-body temperature.
The photodissociation rates for unshielded molecular species, such 
as C$_2$H$_2$, are very much larger than in the other sources 
considered in this study \citep[e.g.][]{RW89}. As a result
C$_2$H$_2$ can never attain significant abundances.
Of course, other factors should be taken into account - such as 
possible variations in the elemental abundances in the ejecta - but 
the extreme clumping required is incompatible with the observed 
spherical coverage of the optically thick dust shell in these objects.
In this case, it seems that other (non PAH-based) formation channels 
must be operating - involving species that can survive the intense 
radation field (e.g. C$_2$, which is partly shielded by the carbon 
ionization continuum \citet{RW89}).\smallskip

CH subgiants have physical characteristics that are closer to Main Sequence 
stars than stars in the Giant phase. They have lower luminosities than 
equivalent stars on the Main Sequence and are known to be chemically peculiar 
\citep{SCL93}, having a carbon over-abundance. The winds from these stars 
are not expected to have the same shock-generated density enhancements as in 
the AGB outflows. 
Thus, even with the extreme parameters we have chosen, we only obtain a 
marginally positive result - with moderate C$_2$H$_2$ abundances obtained
at the lowest photospheric temperatures and highest gas densities
(T$<$4500\,K, n$> 10^{11}$ cm$^{-3}$).
Again, an environment that is conducive to C$_2$H$_2$-PAH chemistry may exist 
if significant clumping occurs, but it would seem that alternative dust 
formation pathways are more likely in these environments.\smallskip

Finally, we find that C$_2$H$_2$ abundances that are sufficient to drive a 
PAH chemistry can be obtained in the winds of R Coronae Borealis stars, 
provided the density is high enough ($\geq 1-5\times10^{11}$ cm$^{-3}$). 
This suggests that the large grains detected in the dust shells of these 
objects \citep{J12} could be produced via a C$_2$H$_2$-PAH chemistry 
operating in dense ejected clouds. By contrast, it is possible that the 
smaller grains observed in these sources may form via an alternative 
mechanism. \smallskip

In summary, we find that C$_2$H$_2$ (and PAH) formation is at least viable in
AGB outflows and the denser parts of R Cor Bor winds, but is only possible in 
the winds of CH subginats if significant clumping is present, and is not 
possible at all in nova winds.
Therefore, on the basis of these studies, it is evident that the formation 
pathway for (carbon) dust cannot be the same in all environments; in particular,
the modified flame-PAH chemsitries that provide a plausible formation channel 
in the dust-rich winds of thermally pulsing AGB stars cannot be applied to
environments such as novae and may only have a marginal importance in RCor Bor 
winds. It would therefore seem that multiple (and, as yet, largely undefined) 
pathways for the kinetic formation of carbon dust exist. 

\section*{Acknowledgements}

HD acknowledges the financial support of the Science and Technology Facilities
Council through a postgraduate studentship.

\bibliographystyle{mn2e}

\label{lastpage}

\end{document}